# APLIKASI BELAJAR MEMBACA IQRO' BERBASIS MOBILE

**Muhammad Sobri[1) ] dan Leon Andretti Abdillah[2)]**

[1)] Manajemen Informatika, Fakultas Ilmu Komputer Universitas Bina Darma
[2)] Sistem Informasi, Fakultas Ilmu Komputer Universitas Bina Darma
Jalan Jenderal Ahmad Yani No.12, Palembang 30264
email : sobri@mail.binadarma.ac.id[1)], leon.abdillah@yahoo.com[2)]

**Abstrak**

*Kemajuan Ilmu Pengetahuan dan Teknologi (IPTEK) hendaklah diikuti dengan Iman dan Takwa (IMTAK), sangat disayangkan, jika masih banyak masyarakat yang belum bisa membaca huruf hijaiyah yang merupakan dasar dari Al-Qur'an. Masyarakat sekarang telah disibukkan dengan berbagai aktipitas sehingga mereka sulit untuk belajar dengan Ustad atau Ustadza mengenai huruf hijaiyah. Untuk mengatasi masalah tersebut maka penulis membuat aplikasi menggunakan Pocket PC karena mudah digunakan, mudah dibawa dan memudahkan masyarakat dalam belajar huruf hijaiyah. Pembuatan aplikasi ini penulis menggunakan metode waterfall yang terdiri dari rekayasa system, analsis kebutuhan perangkat lunak, peranmcangan, pemrograman, pengujian, dan pemeliharaan, sedangkan bahasa pemrogramannya menggunakan Microsoft Visual Basic.NET. Pertama perlu dipersiapkan gambar-gambar huruf hijaiyah dari iqro'1 sampai dengan iqro'4 dengan menggunakan bantuan Photoshop. Selanjutnya perlu pula disiapkan suara dari masing –masing huruf tersebut. Gambar dan suara tersebut disatukan kedalam form-form terkait. Sedangkan code dimasukkan kedalam tiap gambar melalui event click.Hasil dari peneletian ini berupa aplikasi mobile iqro' digital yang dilengkapi dengan suara disetiap gambarnya.*

**Kata kunci :**
*Huruf Hijaiyah, VB .NET, Pocket PC*

## 1. Pendahuluan

Sekarang ini perkembangan komputer sangat pesat seperti adanya *laptop* dan munculnya *pocket pc*. *Pocket PC* merupakan *smart device* PDA yang berjalan menggunakan system operasi *Windows Mobile* buatan *Microsoft*. Akses informasi yang sekarang ini dilakukan dengan komputer atau *laptop* misalnya *email, browsing, chatting, download, internet* dan masih banyak lagi.

Selain itu juga dapat dilakukan dengan menggunakan *pocket pc*, karena telah dilengkapi fasilitas-fasilitas sama halnya dengan komputer atau *laptop*. Bentuk *pocket pc* yang *fleksibel* seperti buku Iqro' memudahkan masyarakat untuk menggunakannya sama halnya seperti menggunakan *handphone*.

Penulis ingin memanfaatkan teknologi tersebut dengan membuat aplikasi huruf hijaiyah, karena selain menggunakan buku iqro' yang telah biasa kita gunakan selama ini untuk belajar huruf hijaiyah.

Aplikasi huruf hijaiyah yang ada pada *pocket pc* ini sangat membantu bagi masyarakat yang mempunyai aktifitas cukup tinggi, jika tidak sempat pergi ke rumah ustad atau ustadza untuk belajar membaca Al-Qur'an terutama huruf hijaiyah. Dengan adanya aplikasi ini memberikan nuansa belajar berbeda kepada masyarakat dalam belajar membaca huruf hijaiyah selain menggunakan buku iqro'.

Dengan adanya aplikasi ini diharapkan nantinya akan sangat membantu bagi masyarakat yang mempunyai cukup banyak kesibukan, sehingga tidak sempat untuk mempelajari iqro' pada ustad atau ustadza secara langsung.

Pengguna aplikasi ini juga dapat dimanfaatkan oleh anak usia dini untuk memperkenalkan teknologi informasi yang bernilai positif bagi pendidikan agama.

## 2. Tinjauan Pustaka

2.1 Aplikasi

Aplikasi berasal dari kata application yang artinya penerapan, lamaran, penggunaan.Secara istilah aplikasi adalah: program siap pakai yang direka untuk melaksanakan suatu fungsi bagi pengguna atau aplikasi yang lain dan dapat digunakan oleh sasaran yang dituju.(www.totalinfo.or.id) [5].

2.2 Pocket PC

*Pocket PC* adalah sebuah smart phone PDA yang berjalan menggunakan sistem operasi *Windows CE* (*CE* sendiri tidak memiliki arti resmi, sering juga disebut sebagai *Compact Edition* atau *Consumer Electronics*) buatan *Microsoft*, walaupun bisa juga menggunakan sistem operasi lain seperti *NetBSD* ataupun Linux. Menurut penciptanya, *Pocket PC* adalah "sebuah alat yang kecil dapat dapat digenggam yang mampu menyimpan, menerima *email, contacts, appointments, tasks,* memainkan file-file multimedia, *games,* menggunakan *Windows Live Messenger (MSN Messenger), surfing* di Internet, dan masih banyak lagi selayaknya *PC desktop*" [6].

2.3 Penelitian Terdahulu

2.3.1 Aplikasi Game Berbasis Flash Untuk Pembelajaran Bagi Anak-Anak

Aplikasi ini memberikan pengajaran kepada anak-anak untuk memudahkan belajar membaca Al-Qur'an dengan cara media permainan berbasis multimedia yang dilengkapi oleh suara, animasi dan gambar [3].

2.3.2 Perancangan Media Pembelajaran "Cara Cepat Belajar Membaca AL-QURAN" Studi Kasus di SD MUHAMMADIYAH CONDONG CATUR YOGYAKARTA

Penelitian ini memberikan pengajaran kepada kepada anak-anak SD yang belajar membaca AL-Qur'an, dengan adanya aplikasi ini memberikan alternative yang lain selain menggunakan buku iqro' yang sifatnya manual [4].

## 3. Metode Penelitian

3.1 Waktu Penelitian

Waktu penelitian dilakukan hampir selama 4 ( empat ) bulan lamanya, terhitung dari tanggal 1 Juli 2012 sampai dengan tanggal 31 Oktober 2012.

3.2 Objek Penelitian

Objek yang penulis gunakan dalam penelitian ini adalah buku Iqro' yang ditulis oleh KH. AS'AD HUMAM dari Balai Litbang LPTQ Nasional, yang diterbitkan oleh Team Tadarus AMN Yogyakarta.

3.3 Metode Pengumpulan Data

Dalam pengumpulan data penulis menggunakan berbagai metode diantaranya adalah: 1) Metode observasi. Yaitu dengan melakukan pengamatan dan pencatatan data yang akan digunakan dalam penelitian ini yaitu buku iqro'. 2) Metode Studi Pustaka, yaitu Studi pustaka, mempelajari, mencari dan mengumpulkan data yang berhubungan dengan penelitian seperti buku dan *internet* yang berkaitan dengan objek permasalahan.

3.4 Metode Pengembangan Sistem

Adapun metode pengembangan sistem yang digunakan dalam pembuatan aplikasi ini adalah tahapan SDLC model waterfall [1].

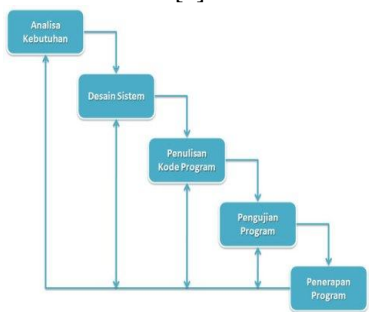

*Gambar.1 Metode Waterfall*

3.5 Alur Pembuatan Aplikasi

Pertama perlu dipersiapkan gambar-gambar huruf hijaiyah dari iqro'1 sampai dengan iqro'4 dengan menggunakan bantuan *Photoshop*. Selanjutnya perlu pula disiapkan suara dari masing –masing huruf tersebut. Gambar dan suara tersebut disatukan kedalam *form-form* terkait. Sedangkan *code* dimasukkan kedalam tiap gambar melalui *event click*.

## 4. Hasil dan Pembahasan

Untuk memastikan agar setiap bagian berjalan dengan baik, maka dilakukan pengujian dengan pendekatan *top-down* [2] (Gambar 2). Aplikasi dibagi menjadi 4 modul utama (F, V, M, T). Pengujian dilakukan dimulai dari sisi sebelah kiri (kelompok F). Apabila sub modul telah berjalan dengan baik maka pengujian berpindah kearah kanan, sampai dengan kelompok T.

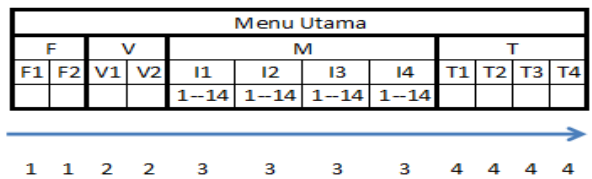

*Gambar.2 Metode Pengujian Top-Down*

Pada pengujian modul kelompok M, perlu diperhatikan sinkronisasi antara huruf hijaiyah dengan suara yang disisipkan.

Setelah semua modul melewati fase pengujian, didapatilah paket aplikasi yang berjalan dengan sempurna. Tampilan awal dari aplikasi ini lihat pada gambar 3.

Aplikasi ini mempunyai *output* berupa suara, jika salah satu gambar huruf hijaiyah diklik maka akan menimbulkan suara bedasarkan gambar yang diklik tersebut.

Penulis berusaha untuk memberikan penjelasan beberapa hasil yang di dapat dalam aplikasi belajar membaca iqro' adapun pembahasan yang dimulai dari pembahasan menu utama, sub menu *file*, sub menu *view*, sub menu materi, sub menu *test* dan sub menu *about*.

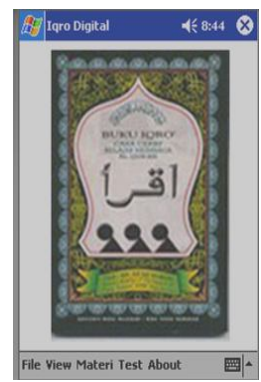

Gambar.3 Menu Utama

4.1 Menu Utama

Menu utama ini terdapat beberapa sub menu antara lain : sub menu file, sub menu view, sub menu materi, sub menu test dan sub menu about. Menu utama berfungsi sebagai pusat aplikasi ini, dimana semua sub menu terdapat pada menu utama ini (Gambar 3).

4.2 Sub Menu *File*

Sub menu *file* berisikan informasi tentang semua huruf hijaiyah beserta cara pengucapannya dan terdapat juga menu exit yang berguna untuk keluar dari aplikasi ini.

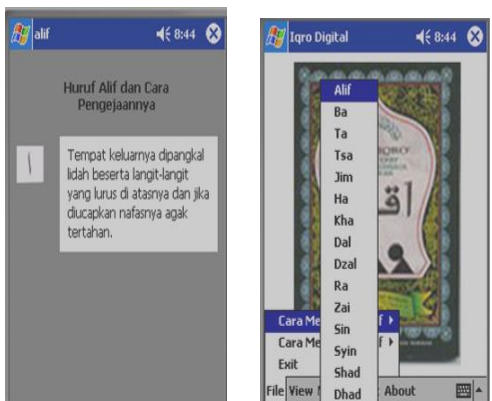
*Gambar.4 Sub Menu File*

4.3 Sub Menu *View*

Sub menu *view* terdapat dua menu yaitu menu cara menggunakannya dan menu mengenal huruf hijaiyah. Menu cara menggunakannya berfungsi untuk memberikan informasi kepada pengguna tentang cara menggunakan aplikasi ini, sedangkan menu mengenal huruf hijaiyah berfungsi untuk memberikan informasi tentang macam-macam huruf hijaiyah.

4.3.1 Sub Menu Cara Menggunakannya

Adapun langkah-langkah untuk membuka sub menu cara menggunakannya adalah dengan cara mngklik sub menu *view,* kemudian pilih menu cara menggunakannya.

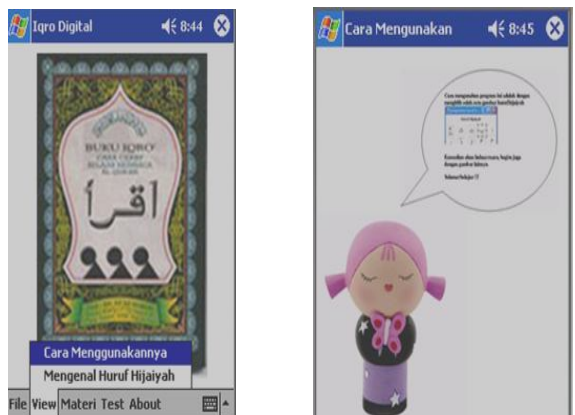
*Gambar.5 Sub Menu Cara Menggunakannya*

4.3.2 Sub Menu Mengenal Huruf Hijaiyah

Adapun langkah-langkah untuk membuka sub menu mengenal huruf hijaiyah adalah dengan cara mngklik sub menu *view,* kemudian pilih menu mengenal huruf hijaiyah.

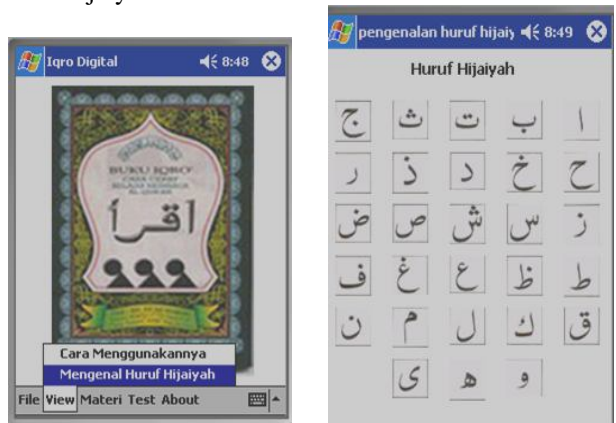
*Gambar.6 Sub Menu Mengenal Huruf Hijaiyah*

4.4 Sub Menu Materi

Sub menu materi memberikan informasi berupa materi pembelajaran dari iqro'1 sampai iqro'4 yang konsepnya seperti buku iqro' yang pada umumnya digunakan.

Di laporan ini penulis hanya menampilkan materi 1 yang terdapat pada Iqro 1, untuk cara menggunakan atau mencoba materi yang lain juga sama cara mengaksesnya.

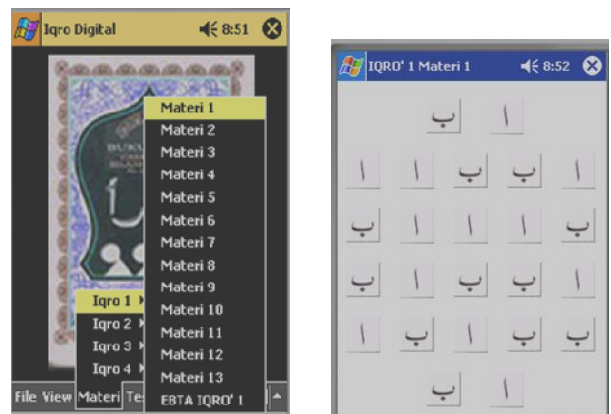
*Gambar.7 Sub Menu Materi*

4.5 Sub Menu *Test*

Sub menu *test* ini, terdapat berbagai test yang berisikan sebuah pertanyaan yang jawabannya *multiple choice,* disini penulis memberikan pertanyaan bedasarkan sub menu materi, sub menu test ini berguna untuk mengetahui apakah pengguna sudah menguasai materi atau belum?

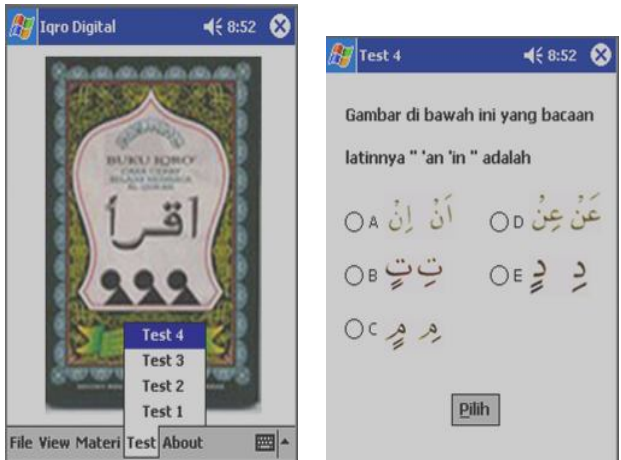

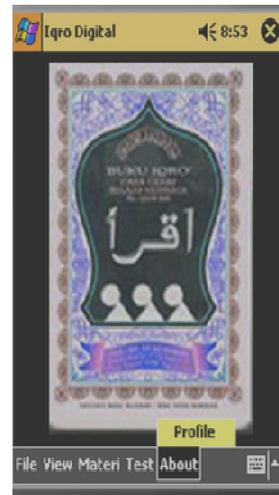

Gambar.8 Sub Menu *Test*

Gambar.11 Menu *About*

Jika jawaban atau pilihan pengguna tepat, aplikasi ini akan memberikan atau menampilkan informasi seperti pada gambar.8 dan sebaliknya jika pengguna jawabannya salah, aplikasi ini akan menampilkan juga informasi seperti gambar.9

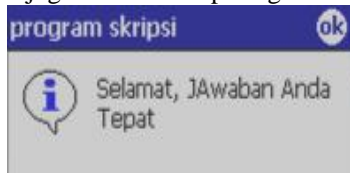

*Gambar.9 Jawaban Anda Benar*

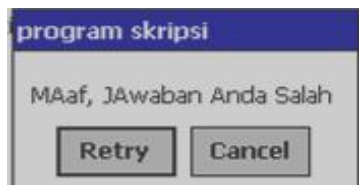

*Gambar.10 Jawaban Anda Salah*

4.6    Sub Menu *About*

Sub menu about ini berisikan informasi – informasi mengenai aplikasi ini dan informasi profile pembuat aplikasinya .kepada pengguna aplikasi ini.

4.7    Kelebihan dan Kekurangannya
Penulis menyadari sepenuhnya bahwa aplikasi yang penulis buat ini mempunyai beberapa kelebihan dan beberapa kekuranganya sebagai berikut :

**4.7.1    Kelebihannya**
Adapun kelebihan aplikasi ini antara lain :
1) Memudahkan orang awam untuk pelajar huruf hijaiyah karena disertai suara pada gambar huruf hijaiyah, jika salah satu gambar huruf hijaiyah diklik akan mengeluarkan suara bedasarkan gambar yang diklik tersebut.
2) Memudahkan masyarakat belajar huruf hijaiyah selain menggunakan buku iqro.
3) Dapat belajar dimanapun, karena sifatnya *fleksibel* seperti menggunakan *handphone*.

**4.7.2    Kekurangannya**
Adapun kekurangan program ini antara lain:
1) Program ini pembelajarannya sampai iqro' 4.
2) Tampilanya masih sangat sederhana.

## 5. Kesimpulan dan Saran

5.1    Kesimpulan
Berdasarkan dari pembuatan aplikasi belajar membaca iqro' tersebut, maka penulis dapat menarik kesimpulan sebagai berikut : 1) Menghasilkan sebuah aplikasi pembelajaran iqro' yang berjalan pada *smart device( mobile* yang *fleksibel* seperti buku Iqro' semudah menggunakan *handphone*. 2) Dengan adanya aplikasi ini memberikan nuansa belajar yang berbeda pada pembelajaran huruf hijaiyah pada umumnya seperti buku iqro'.

5.2    Saran
Berdasarkan dari pembuatan aplikasi belajar membaca iqro' tersebut, maka penulis memberikan saran sebagai berikut : 1)Aplikasi yang telah dihasilkan hanya bisa berjalan pada smart device ( *pocket pc* ) atau PDA ( *Personal Digital Asistan* ) sehingga perluh dikembangkan agar bisa digunakan pada semua jenis *handphone*. 2) Aplikasi dibuat sampai dengan Iqro' 4 sehingga perluh dikembangkan sampai dengan Iqro' 6.

## Daftar Pustaka

## Biodata Penulis

**Muhammad Sobri**, memperoleh gelar Sarjana komputer (S.kom), Program Studi Teknik Informatika Univeristas Bina Darma (UBD), lulus tahun 2009. Tahun 2011 memperoleh gelar Magister Komputer (M.Kom) dari Konsentrasi Software Engineering UBD. Saat ini sebagai Staf Pengajar program studi Manjemen Informatika UBD Palembang.

**Leon Andretti Abdillah**, mendapatkan gelar Sarjana Ilmu Komputer, pada Program studi Sistm Informasi dari STMIK Bina Darma tahun 2001, dan Magister Managemen, Konsentrasi Sistem Informasi dari Universitas Bina Darma tahun 2006. Beliau pernah melanjutkan studi PhD di The University of Adelaide (2010-2012) pada School of Computer Science. Saat ini, beliau bekerja sebagai dosen pada Universitas Bina Darma pada Program Studi Sistem Informasi. Minat riset utama beliau adalah *Information Systems, Scientific Journals, Information Retrieval, Human Resource MIS, Database Systems, Programming,* dan *Entrepreneur*.